\documentclass[a4paper, english, final, amsfonts, amssymb, amsmath, reprint, showkeys, nofootinbib, twoside, superscriptaddress]{revtex4-1}
\usepackage[english]{babel}
\usepackage[utf8]{inputenc}
\usepackage[colorinlistoftodos, color=green!40, prependcaption]{todonotes}
\usepackage{amsthm}
\usepackage{mathtools}
\usepackage{physics}
\usepackage{xcolor}
\usepackage{graphicx}
\usepackage[left=23mm,right=13mm,top=35mm,columnsep=15pt]{geometry} 
\usepackage{adjustbox}
\usepackage{placeins}
\usepackage[T1]{fontenc}
\usepackage{lipsum}
\usepackage{csquotes}
\usepackage[pdftex, pdftitle={Article}, pdfauthor={Author}]{hyperref} 
\begin{document}
\title{Optimising microstructural characterisation of white-matter phantoms: impact of gradient waveform modulation on Non-uniform Oscillating Gradient Spin-Echo sequences}

\author{Melisa L. Gimenez}
    \email[Correspondence email address: ]{melisa.gimenez@ib.edu.ar}
    \affiliation{Centro At\'omico Bariloche, CNEA, S.C. de Bariloche, Argentina}
    \affiliation{Instituto Balseiro, CNEA, Universidad Nacional de Cuyo, S.C. de Bariloche, Argentina}
\author{Pablo Jimenez}
    \affiliation{Instituto Balseiro, CNEA, Universidad Nacional de Cuyo, S.C. de Bariloche, Argentina}
    \affiliation{Centro At\'omico Bariloche, CONICET, CNEA, S.C. de Bariloche, Argentina}
\author{Leonardo A. Pedraza P\'erez}
    \affiliation{Centro At\'omico Bariloche, CNEA, S.C. de Bariloche, Argentina}
    \affiliation{Instituto Balseiro, CNEA, Universidad Nacional de Cuyo, S.C. de Bariloche, Argentina}
\author{Diana Betancourth}
    \affiliation{Centro At\'omico Bariloche, CONICET, CNEA, S.C. de Bariloche, Argentina}
\author{Analia Zwick}
    \affiliation{Instituto Balseiro, CNEA, Universidad Nacional de Cuyo, S.C. de Bariloche, Argentina}
    \affiliation{Centro At\'omico Bariloche, CONICET, CNEA, S.C. de Bariloche, Argentina}
    \affiliation{Instituto de Nanociencia y Nanotecnología, CNEA, CONICET, S.C. de Bariloche, Argentina}
\author{Gonzalo A \'Alvarez}
    \email[Correspondence email address: ]{gonzalo.alvarez@conicet.gov.ar}
    \affiliation{Instituto Balseiro, CNEA, Universidad Nacional de Cuyo, S.C. de Bariloche, Argentina}
    \affiliation{Centro At\'omico Bariloche, CONICET, CNEA, S.C. de Bariloche, Argentina}
    \affiliation{Instituto de Nanociencia y Nanotecnología, CNEA, CONICET, S.C. de Bariloche, Argentina}


\begin{abstract}
Changes in the nervous system due to neurological diseases take place at very small spatial scales, on the order of the micro and nanometers. Developing non-invasive imaging methods for obtaining this microscopic information as quantitative biomarkers is therefore crucial for improved medical diagnosis. In this context, diffusion weighted magnetic resonance imaging has shown significant advances in revealing tissue microstructural features by probing molecular diffusion processes. Implementing modulated gradient spin-echo sequences allows monitoring time-dependent diffusion processes to reveal such detailed information. In particular, one of those sequences termed Non-uniform Oscillating Gradient Spin-Echo (NOGSE), has shown to selectively characterise microstructure sizes by generating an image contrast based on a signal decay-shift rather than on the conventionally used signal decay rate. In this work, we prove that such decay-shift is more pronounced with instantaneous switches of the sign of the magnetic field gradient strength. As fast gradient ramps need to be avoided in clinical settings, due to potential patient discomfort and artefacts in imaging, we evaluate the method's efficacy for estimating microstructure sizes using both idealised, sharp gradient modulations and more realistic, smooth modulations. In this more realistic scenario we find that the signal decay-shift might be lost as the diffusion time increases, likely hindering the accurate estimation of microstructural characteristics. We demonstrate, by a combination of numerical simulations, information theory analysis and proof-of-principle experiments with white-matter phantoms, that optimal sequence design to estimate microstructure size distributions can be achieved using either sharp or smooth gradient spin-echo modulations. This approach simplifies the translation of the NOGSE method for its use in clinical settings.
\end{abstract}


\maketitle

\renewcommand\thefootnote{\fnsymbol{footnote}}
\setcounter{footnote}{1}

\section{Introduction}
\label{Introduction}
Magnetic resonance imaging (MRI) has become an indispensable tool in biomedicine and in clinical diagnosis as it provides valuable insights into the human body with non-invasive imaging techniques~\cite{desantis_nongaussian_2011,white_probing_2013,white_diffusion-weighted_2014,alexander_imaging_2019,xu_magnetic_2020,fokkinga_advanced_2023}. The avoidance of invasive approaches holds particular significance, especially in the field of brain imaging. One of the most challenging goals of MRI is to characterise microstructural features of tissues in the central and peripheral nervous system~\cite{le_bihan_looking_2003,xu_mapping_2014,alexander_imaging_2019}. The estimation of diameters, density and orientations of axons and myelin sheath thickness are essential for understanding the biophysical mechanisms behind several neurodegenerative diseases such as Alzheimer's~\cite{Jack_Alzheimer_2008,drago_disease_2011,harrison_imaging_2020,chu_comparison_2022}, autism~\cite{Hyde_Neuroanatomical_2010}, amyotrophic lateral sclerosis~\cite{enzinger_nonconventional_2015,grussu_neurite_2017}, and schizophrenia~\cite{kubicki_review_2007}; for detecting ischaemia in real time~\cite{canazza_experimental_2014,zhu_diffusion_2019}; for tractography~\cite{caminiti_diameter_2013,maier_challenge_2017,schilling_prevalence_2022} and for pre-surgical planning~\cite{maffei_diffusion-based_2019,maffei_missing_2019,huang_brain_2022}.

Due to the great potential of MRI, there is a tireless endeavour to derive trustable and quantitative imaging biomarkers sensitive to microscopic changes that affect tissue functionality related to a wide range of pathologies~\cite{assaf_composite_2005,assaf_axcaliber_2008,patterson_technology_2008,padhani_diffusion-weighted_2009,ong_quantifying_2010,grussu_neurite_2017,jelescu_challenges_2020}. Diffusion-weighted MRI (DWI) is a promising approach to account for it as DWI senses the diffusion of water molecules within tissues to infer microstructural boundaries, such as those imparted by cellular membranes~\cite{le_bihan_looking_2003,kakkar_low_2018,alexander_imaging_2019,veraart_noninvasive_2020,olesen_diffison_2022}. Quantitative approaches provide microstructure parameters that may play the role of quantitative histological or physiological markers~\cite{assaf_composite_2005,assaf_axcaliber_2008,alexander_orientationally_2010,zhang_noddi_2012,shemesh_measuring_2013,novikov_quantifying_2019,palombo_SANDI_2020,capiglioni_noninvasive_2021,warner_temporal_2023}.

Despite the widespread use of DWI, the millimetre-scale resolution of clinical MRI continues to present a difficulty in reliably estimating tissue properties at the cellular level of a few micrometres~\cite{jelescu_design_2017,novikov_modeling_2018,afzali_sensitivity_2021}. The key challenge to overcome then, is the development of robust methods with reproducible results, enabling the inference of those mesoscopic tissue characteristics by analysing the averaged signal arising from each elementary volume of the image (voxel).

Among the various DWI methods, Modulated Gradient Spin-Echo (MGSE) sequences are emerging as a promising paradigm for extracting this information. They achieve it by probing spatial scales of the molecular motion by suitably modulating the frequencies of the magnetic field gradients~\cite{stepisnik_time-dependent_1993,callaghan_frequency-domain_1995,stepisnik_spectral_2006,callaghan_translational_2011,siow_estimation_2012,shemesh_measuring_2013,alvarez_coherent_2013,sjolund_constrained_2015,drobnjak_pgse_2016,nilsson_resolution_2017}. These sequences, such as the Oscillating Gradient Spin Echo (OGSE) sequences~\cite{stepivsnik_analysis_1981,stepisnik_time-dependent_1993,callaghan_frequency-domain_1995, does_oscillating_2003,gore_characterization_2010}, are a generalisation of the Pulse Gradient Spin-Echo (PGSE) sequence~\cite{stejskal_spin_1965} and can generate high-quality microstructural images. The conventional DWI approach under MGSE sequences typically relies on extracting apparent diffusion coefficients from the rate of the magnetisation signal decay~\cite{stejskal_spin_1965,le_bihan_looking_2003,grebenkov_nmr_2007,callaghan_translational_2011}.

Recent developments inspired on quantum information tools~\cite{smith_shift_2012,zwick_quantum_2023} have shown that non-uniform MGSE sequences can selectively monitor molecular diffusion processes and, thereby, probe tissue microstructure sizes with the quantitative information encoded directly in a signal decay-shift rather than in the signal decay-rate~\cite{shemesh_measuring_2013,alvarez_coherent_2013}. This decay-shift can be exploited to selectively probe microstructure sizes with a non-uniform oscillating gradient spin-echo (NOGSE) sequence~\cite{shemesh_measuring_2013,alvarez_coherent_2013,shemesh_size_2015,capiglioni_noninvasive_2021}, a variant of an OGSE sequence~\cite{callaghan_frequency-domain_1995,stepisnik_spectral_2006,gore_characterization_2010,drobnjak_pgse_2016} that consists of concatenating two modulated gradient sequences with different frequencies.

The decay-shift in the NOGSE signal provides a higher parametric sensitivity to compartment size changes compared to the one gained from the decay rate, while it factors out experimental imperfections and relaxation weightings~\cite{shemesh_measuring_2013,alvarez_coherent_2013}. As shown in this work, this effect ideally arises with sharp magnetic field gradient modulations, \textit{i.e.} when using instantaneous switching of the gradient's sign.

In this study, we pursue two primary objectives: firstly, to assess the reliability and self-consistency of compartment-size distribution estimation using NOGSE and, secondly, to compare the effectiveness of this inference between idealised-sharp and realistic-smooth gradient modulation techniques, with the latter being more conducive to clinical applications.

Our investigation delves into the parametric sensitivity of NOGSE under both sharp and smooth modulations, leveraging a combination of numerical simulations and information theory analysis. Our findings reveal that optimal estimation of microstructure size distributions is achievable regardless of the gradient modulation type. We bolster these results through proof-of-principle experiments employing phantoms that replicate typical white-matter restriction-size distributions, showcasing the method's reliability and self-consistency in agreement with our numerical and analytical assessments. This study thus helps pave the way towards the development of diagnostic techniques based on quantitative imaging, furnishing valuable tools for estimating microstructural features beyond the resolution limits of DWI voxel size.

\section{Theory}
\label{Theory}
\subsection{Diffusion-weighted MRI signal}
\label{DWMR}
We focus on extracting information regarding the microstructure sizes within tissues or porous media by monitoring the diffusion of water molecules in confined compartments. The nuclear magnetic resonance (NMR) signal is affected by the dephasing effects due to the Brownian motion of the molecules within these compartments.

The NMR signal evolution is determined from the interaction of the ensemble of 1H nuclear $1/2$-spins with an external static magnetic field $B_0$ and an applied magnetic field gradient $G(t)$ along a spatial direction $\hat{r}$~\cite{grebenkov_nmr_2007,callaghan_translational_2011}. The gradient strength $G(t)$ is modulated as a function of time as $G(t)$\,=\,$G f(t)$, varying its sign and strength with the function $\left| f(t)\right|\leq1$~\cite{stepisnik_time-dependent_1993,callaghan_frequency-domain_1995,gore_characterization_2010}. Here, $G$ is the gradient strength. The static field $B_0$ defines the Larmor frequency $\gamma B_0$, with $\gamma$ the gyromagnetic ratio of the nucleus. In the rotating frame of reference at such frequency, the spin's precession frequency $\omega(t)$\,=\,$\gamma G(t) r(t)$ is determined by the instantaneous gradient strength and position of the spin, thus fluctuating due to diffusion-induced displacements of the water molecules~\cite{grebenkov_nmr_2007,callaghan_translational_2011}. The spin position $r(t)$ describes the diffusing trajectory of the nuclear spin along the field gradient direction.

The spins acquire a phase $\phi(t_D)$ according to their random displacements as a function of the total diffusion time $t_D$. The random phase $\phi(t_D)$ typically follows a Gaussian phase distribution~\cite{klauder_spectral_1962,stepisnik_analysis_1981,stepisnik_validity_1999}. Hence, the diffusion-weighted NMR signal becomes
\begin{equation}
	M(t_D)=M(0) e^{-\beta(t_D)} = M(0) e^{-\frac{1}{2}\langle \phi^2(t_D)\rangle},
	\label{Eq_Magn}
\end{equation}
which is determined from the spin ensemble average. It decays as a function of time, depending on the random phase's variance $\langle \phi^2(t_D)\rangle$ over the spin ensemble.

The attenuation factor $\beta(t_D)$ can be written in a Fourier representation as~\cite{stepisnik_time-dependent_1993,lasic_displacement_2006,shemesh_measuring_2013,alvarez_coherent_2013,zwick_maximizing_2016}
\begin{equation}
	\beta(t_D)=\frac{\gamma^2}{2}\int_{-\infty}^{\infty}\left|  F(\omega,t_D) \right|^2 S(\omega) d\omega,  
	\label{Eq_beta}
\end{equation}
where $F(\omega,t_D)$ is a filter function defined by the Fourier Transform (FT) of the gradient modulation $G(t)$ and $S(\omega)$ is the displacement spectral density given by the FT of the spin's displacement correlation function $\langle \Delta r(0)\Delta r(t) \rangle$. The spin's displacement $\Delta r(t)$ is given by $r(t)-\langle r(t)\rangle$. Within this FT representation, the signal decay is determined by the integral of the overlap between the filter function $F(\omega,t_D)$ and the spectral density $S(\omega)$. The filter $F(\omega,t_D)$ depends exclusively on the setup parameters of the chosen MGSE sequence, while the spectral density $S(\omega)$ depends on the characteristics of the diffusion-driven fluctuations. We consider that in typical restricted diffusion conditions, it is well approximated by a Lorentzian function~\cite{stepisnik_time-dependent_1993,shemesh_measuring_2013,alvarez_coherent_2013,zwick_precision_2020}
\begin{equation}
	S(\omega)= \frac{D_0 \tau_c^2}{(1+\omega^2 \tau_c^2)\pi}.
	\label{Eq_DensidadEspectral}
\end{equation}
The free diffusion coefficient is $D_0$ and $\tau_c$ the time required for most molecules to fully probe the compartment boundaries. Considering the Einstein's expression $l_c^2$\,=\,$2 D_0 \tau_c$, the characteristic time $\tau_c$ is related to the diffusion restriction length $l_c$ imposed by the microstructure compartment in which diffusion is taking place~\cite{callaghan_translational_2011}. 

Therefore, by suitably choosing the oscillating gradient modulation parameters to steer the NMR signal decay, one can probe the water diffusion process and, consequently, extract microstructural restriction sizes from its time dependence~\cite{gore_characterization_2010,shemesh_measuring_2013,alvarez_coherent_2013,xu_mapping_2014,shemesh_size_2015,capiglioni_noninvasive_2021}.

\subsection{Size distribution imaging} 
Morphological heterogeneities in most biological tissues and porous media lead to multiple compartment sizes. To describe these complex systems, a probability distribution for the compartment sizes is typically used~\cite{assaf_axcaliber_2008,barazany_vivo_2009,nilsson_role_2013}. In the particular case of white-matter tissue, a lognormal probability distribution $P(l)$ is often observed for the restriction sizes $l$~\cite{pajevic_optimum_2013,liewald_distribution_2014,shemesh_size_2015,capiglioni_noninvasive_2021}. Hence, we consider such distribution to describe the interstitial compartment sizes on the white-matter phantoms used in this work, which is given by
\begin{equation}
	P(l) = \frac{1}{\sigma\sqrt{2\pi} l}e^{-\frac{ \left(  \ln{l} - \mu \right)^2 }{2\sigma^2 }},
	\label{Eq_LogNormal}
\end{equation}
with the mean size $l_c$\,=\,$e^{\mu+\sigma^2/2}$, the standard deviation $\sigma_c$\,=\,$\sqrt{e^{ \sigma^2+2\mu} \left( e^{\sigma^2}-1 \right)}$, the median $e^{\mu}$ and the mode ${e^{\mu -\sigma ^{2}}}$.

The total signal originated from the spins diffusing within this compartment size distribution model is therefore determined by the weighted average
\begin{equation}
	M(t_D)= \frac{\Sigma_l P(l) M_l(t_D)}{\Sigma_l P(l)},
	\label{Eq_MagnDistr}
\end{equation}
where the magnetisation signal $M_l(t_D)$ corresponds to the signal given by spins diffusing within a compartment size $l$. In the implemented fitting model we consider a discrete version of the lognormal distribution. Including the factor $1/ \Sigma_l P(l)$ we ensure the normalisation of the distribution.

\subsection{NOGSE sequence} 
\begin{figure*}[ht]
	\includegraphics[width=1\textwidth]{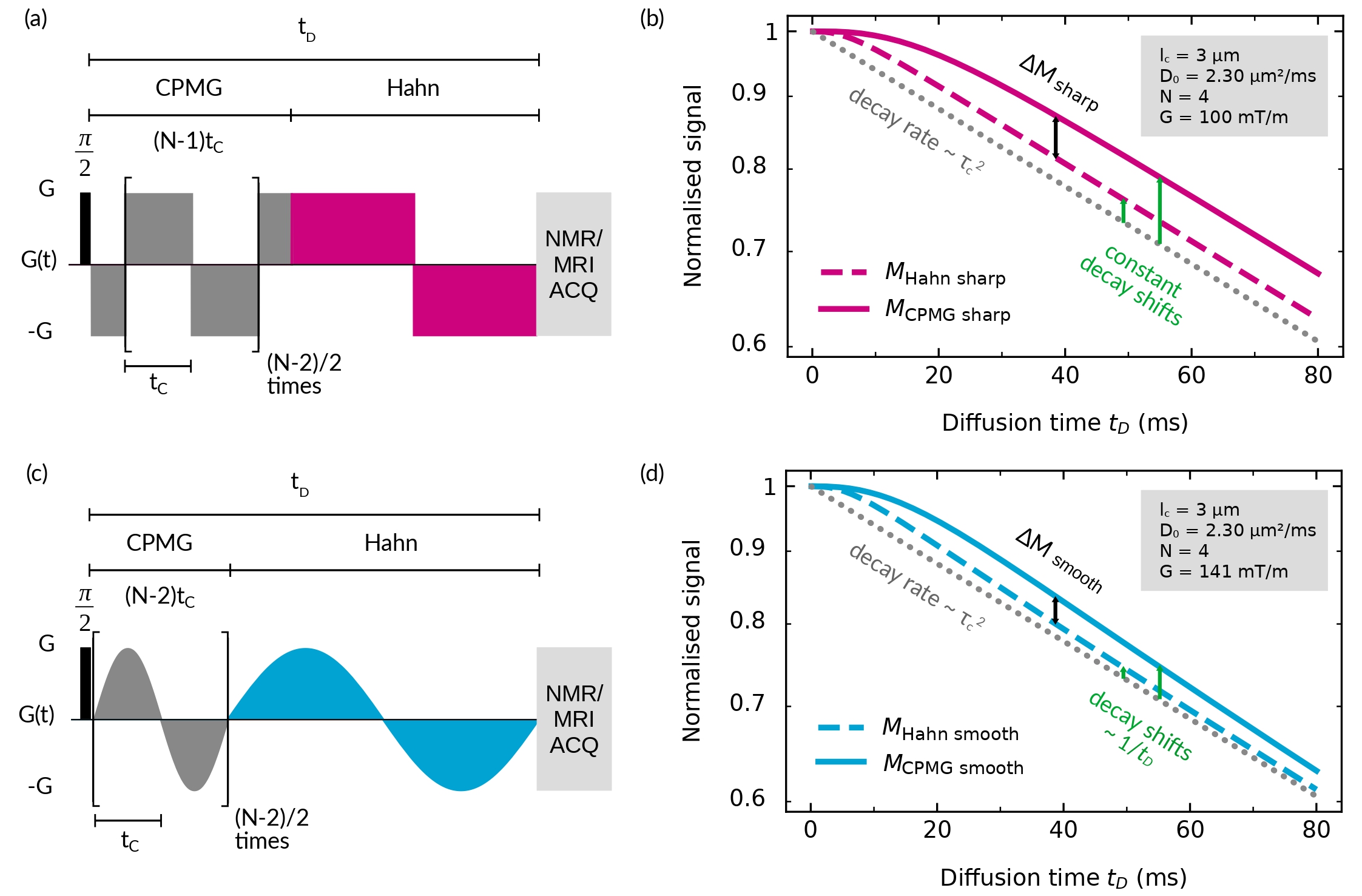}
	\caption{NOGSE sequence and signal decay as a function of the diffusion time $t_D$ for the ideal, sharp (squared) gradient modulation (panels a and b) and for the smooth (sinusoidal) gradient modulation (panels c and d). (a, c) The NOGSE~\cite{shemesh_measuring_2013} sequence, with sharp (a) and smooth (c) gradient modulations, is a variant of an OGSE sequence that consists of concatenating two modulated gradient sequences with different frequencies and factors out experimental imperfections and relaxation weightings. (b, d) Signal decay as a function of the diffusion time $t_D$ for water molecules diffusing within a single restriction size for sharp (b) and smooth (d) gradient modulations. We show the two limiting cases for the NOGSE lobe duration $t_C\rightarrow0$ (Hahn: dashed lines) and $t_C$\,=\,$t_D/N$ (CPMG: continuous lines). The sharp and smooth NOGSE contrasts, $\Delta M_{sharp}$ and $\Delta M_{smooth}$, are sensitive to microstructure sizes. At the restricted diffusion regime $t_C\gg \tau_c$, the signals decay with a constant rate that is independent of the number of lobes and the same for both gradient modulations. The decay-shifts (green arrows) are independent of the total diffusion time $t_D$ for the sharp, instantaneous ramps of the gradient modulation (b). However, if smooth ramps are present the decay-shift decreases with the diffusion time $t_D$~(d).
		\label{Figure_1}}
\end{figure*}

The NOGSE sequence~\cite{shemesh_measuring_2013,alvarez_coherent_2013} is depicted in Figure~\ref{Figure_1}a, schematized with a sharp modulation, \textit{i.e.} with instantaneous switching of the gradient's sign. It comprises two concatenated sequences that produce a signal decay similar to that of a CPMG (OGSE) modulation~\cite{carr_effects_1954,meiboom_modified_1958} followed by a Hahn (Gradient Echo) modulation~\cite{hahn_spin_1950}. The total diffusion time is $t_D$ and the total number of gradient lobes is $N$. For the sharp gradient modulation, refocusing lobes of duration $t_C$ are repeated $N-1$ times, corresponding to the CPMG modulation. Notice that, as the modulation shape is based on a cosine symmetry, the first and last gradient pulses are of duration $t_C/2$. The last refocusing period corresponds to the Hahn counterpart with a total duration $t_D-(N-1)t_C$. As $t_D$, $N$ and the Echo Time between the acquired images are kept fixed, gradient switching imperfections and additional relaxation effects are factored out~\cite{shemesh_measuring_2013,alvarez_coherent_2013}.

As instantaneous switching of the modulating gradients is not experimentally achievable, we performed the experiments with trapezoidal waveforms with the fastest slew rate possible. This sets a paradigmatic sharp gradient modulation sequence, which one can approximate as an instantaneous switch of the gradient sign, if gradient ramp times are much shorter than the lobes duration $t_C$. If not, they are assumed as gradient switching imperfections that are factored out by the NOGSE approach~\cite{shemesh_measuring_2013,alvarez_coherent_2013}.

\subsection{NOGSE decay-shift for sharp modulations} 
The NOGSE sequence selectively probes microstructure compartment sizes from a signal decay-shift, rather than from a signal decay-rate~\cite{shemesh_measuring_2013, alvarez_coherent_2013,shemesh_size_2015,capiglioni_noninvasive_2021}. Figure~\ref{Figure_1}b shows the signal decay of the sharp-modulated NOGSE at the limiting values of the lobe duration $t_C$, $M_{NOGSE}(t_D,t_C\sim 0,N)$\,=\,$M_{Hahn}(t_D)$ for $t_C$\,=\,0 and $M_{NOGSE}(t_D,t_C=t_D/N,N)$\,=\,$M_{CPMG}(t_D,N)$ for $t_C$\,=\,$t_D/N$. Analysing the magnetisation signal as a function of the diffusion time $t_D$, the NOGSE contrast $\Delta M$ is determined from the signal difference between the magnetisation decay at $t_D$ at the limiting values of $t_C$, described above. The difference between these two limiting values is the NOGSE modulation contrast $\Delta M$\,=\,$M_{CPMG}(t_D,N)-M_{Hahn}(t_D)$.

In the restricted diffusion regime, the diffusing spins experience a spatial restriction imposed by the confining cavity. To sense these restrictions, the lobe duration $t_C$ needs to be longer than the restriction correlation time $\tau_c$. At $t_C$\,=\,0 the NOGSE signal decays with the attenuation factor

\begin{equation}
	\beta^{restr}_{Hahn \, sharp}(t_D)=\gamma^2 G^2 D_0 \tau_c^2(t_D-3\tau_c),
	\label{Eq_BetaRestrHahnC}
\end{equation}
while at $t_C=t_D/N$, with
\begin{equation}
	\beta^{restr}_{CPMG \, sharp}(t_D,N)=\gamma^2 G^2 D_0 \tau_c^2 \left( t_D-(2N+1)\tau_c \right).
	\label{Eq_BetaRestrCPMGC}
\end{equation}
Therefore, the NOGSE contrast with a sharp gradient modulation at the diffusion restricted regime is (see Appendix~\ref{app_cuad})~\cite{shemesh_measuring_2013,alvarez_coherent_2013,capiglioni_noninvasive_2021}

\begin{multline}
	\Delta M^{restr}_{sharp}(t_D,N)=\\
	\left( e^{ \gamma^2 G^2 D_0 \tau_c^3 2(N-1) } -1 \right)  \left( e^{-\gamma^2 G^2 D_0 \tau_c^2 \left( t_D - 3\tau_c \right) } \right).    
	\label{Eq_DeltaMrestr_cuad}
\end{multline}
This signal evolution evidences a decay-rate $\gamma^2 G^2 D_0 \tau_c^2$ that is independent of the number of lobes and it is proportional to $\tau_c^2$, as in OGSE-type experiments~\cite{klauder_spectral_1962,stepisnik_time-dependent_1993,callaghan_frequency-domain_1995,baron_oscillating_2014,drobnjak_pgse_2016}. The signal behaviour manifests a decay-shift, which is a constant term $\gamma^2 G^2 D_0 \tau_c^3(1+2N)$ independent of the total diffusion time $t_D$ and varies as $\tau_c^3$~\cite{shemesh_measuring_2013,alvarez_coherent_2013}. This NOGSE decay-shift arises if the gradient modulations are sharp, with a parametric sensitivity to microstructure compartment sizes proportional to $l_c^6$, enabling selective probing of specific restriction sizes~\cite{capiglioni_noninvasive_2021}.

\subsection{NOGSE decay-shift for smooth modulations} 
The NOGSE contrast $\Delta M^{restr}_{sharp}$ of Equation~\ref{Eq_DeltaMrestr_cuad} results from considering ideal gradient pulses, with instantaneous ramps for the switching of the applied gradient amplitude (Figure~\ref{Figure_1}a). We define this case as the sharp gradient modulation that gives a decay-shift independent of the diffusion time $t_D$. However, we observe that if we deviate from the ideal case scenario, \textit{i.e.} if ramps are smooth (Figure~\ref{Figure_1}c), the decay-shift decreases as a function of $t_D$ (see Figure~\ref{Figure_1}d). In this study we consider for smooth gradient modulations the paradigmatic case of a sinusoidal functional dependence. Here, the first $N-2$ lobes, with $N\geq4$, correspond to the CPMG modulation and the last two to the Hahn modulation, with a duration $t_D-(N-2)t_C$.

In order to make both implementations of NOGSE modulations comparable in the scanner, $t_D$, $N$, $t_C$ and $G$ take the same acquisition parameters for both types of gradient modulations.

The attenuation factor for this smooth NOGSE version at the restricted diffusion regime is (see Appendix~\ref{app_SIN})
\begin{equation}
	\beta^{restr}_{Hahn \, smooth}=\frac{\gamma^2 G^2 D_0 \tau_c^2}{2} \left( t_D - \frac{4 \pi^2 \tau_c^2}{t_D} \right)
	\label{Eq_BetaRestrHahnS}
\end{equation}
for the Hahn modulation at the limiting value $t_C\rightarrow0$ and
\begin{equation}
	\beta^{restr}_{CPMG \, smooth}=\frac{\gamma^2 G^2 D_0 \tau_c^2}{2} \left( t_D - \frac{N^2 \pi^2 \tau_c^2}{t_D} \right)
	\label{Eq_BetaRestrCPMGS}
\end{equation}
for the CPMG modulation at the limiting value $t_C=t_D/N$. Therefore, for the smooth gradient modulation, the NOGSE contrast is

\begin{multline}
	\Delta M^{restr}_{smooth}(t_D,N)=\\
	\left( e^{\frac{\pi^2 \gamma^2 G^2 D_0 \tau_c^4 \left( N^2-4 \right) }{2 t_D}} -1 \right)
	\left( e^{-\frac{\gamma^2 G^2 D_0 \tau_c^2}{2} \left( t_D-\frac{4\pi^2\tau_c^2}{t_D} \right) } \right) .      
	\label{Eq_DeltaMrestr_sin}
\end{multline}

Here, the smooth modulation evidences a parametric sensitivity to probe compartment sizes proportional to $l_c^8$. However, the shift decays as a function of the diffusion time as a power law proportional to $1/t_D$. This indicates that at very long diffusion times $t_D \gg \tau_c$, the decay-shift might become non-observable.

\section{Methods}
\label{Experimental}
\subsection{Phantom manufacturing}
\label{Phantom}
\begin{figure*}[ht]
	\centerline{\includegraphics[width=\textwidth]{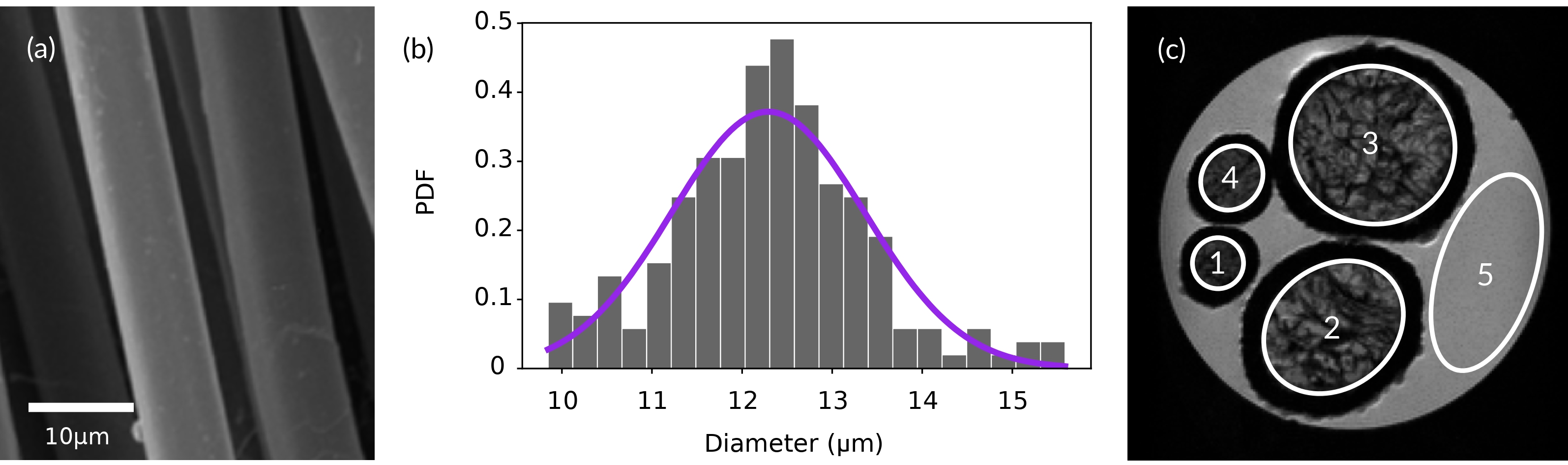}}
	\caption{White-matter phantom using bundles of aramid fibres immersed in water. (a) A SEM image of the aramid fibres. These fibres are used to emulate restriction-size distributions similar to those found in extra-axonal space in the white matter of the human brain. (b) Histogram of the aramid fibre diameters, showing a mean diameter of 12.3\,$\mu$m. (c) Transverse plane dMRI of the phantom with bundles of aramid fibres immersed in water in a 15\,ml plastic tube. The numbered regions represent the different ROIs analysed in this work: ROIs 1 to 4 correspond to aramid fibre bundles with different packing density and ROI5 only contains free water.
		\label{Figure_2}}
\end{figure*}

We mimic axon bundles with phantoms comprising aramid fibres of $\sim$12\,$\mu$m mean diameter. A Scanning electron microscope (SEM) image of the fibres is shown in Figure~\ref{Figure_2}a and a histogram to determine their diameter distribution is shown in Figure~\ref{Figure_2}b. The samples for the SEM were placed on a sample holder and glued on a carbon tape. Subsequently, the sample holder was placed in a gold bath for one minute, generating a gold layer of approximately 10\,nm.

To prepare the phantoms, fibres were aligned together and positioned inside a thermo-shrinkable plastic tube to form a fibre bundle. Then, it was heated to enhance the compaction degree. Several fibre bundles were put into a 15\,ml conical centrifuge tube, and then they were slowly filled with demineralized water so as to let the water enter the fibre bundles by capillary effect. Once the phantom was completely immersed in water, it was left in ultrasound to remove the remaining air bubbles.

As the fibres are both solid and impermeable, the water diffuses within the interstices of the extra-fibre region. The presence of fibres provides a tortuous environment with displacement restrictions for the diffusing water molecules, mimicking the extra-axonal space in white matter tracts.

A transverse plane DWI of the phantom is shown in Figure~\ref{Figure_2}c, where parallel aramid fibre bundles of different packing densities are shown. We define several regions-of-interest (ROIs) on the transversal plane to evaluate the microstructure restriction-size distribution of the interstitial compartments. Outside the fibre bundles in the phantom (ROI5), there is freely diffusing water that is used for calibration, control purposes and to determine the free diffusion coefficient.

\subsection{Image acquisition and data processing}
\begin{figure*}[ht]
	\centerline{\includegraphics[width=\textwidth]{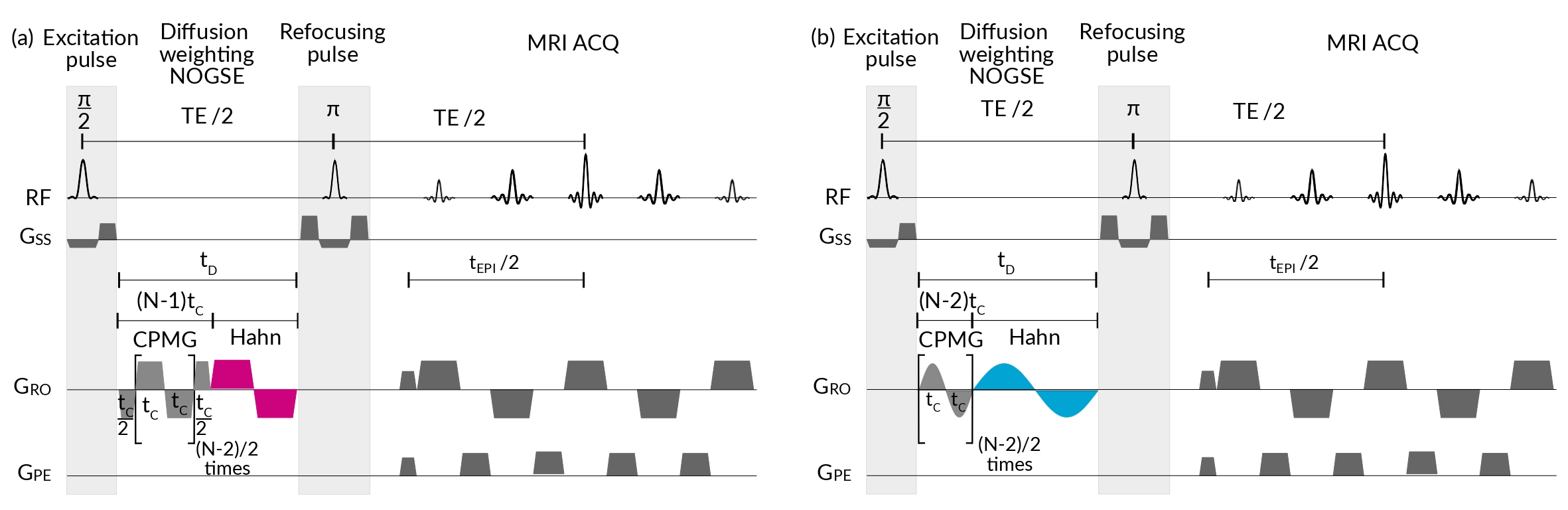}}
	\caption{Scheme of the experimentally implemented NOGSE sequence for (a) sharp and (b) smooth gradient modulations. An initial selective radiofrequency excitation $\pi/2$-pulse is applied for slice selection. It is followed by the diffusion weighting building block with the NOGSE gradient modulations of duration $t_D$, based on the schemes described in Figure~\ref{Figure_1}. The gradient strength and sign are modulated with trapezoidal shapes (a) or sinusoidal shapes (b) during the diffusion time $t_D$. A selective radiofrequency $\pi$-pulse is then applied to refocus magnetic field inhomogeneities. Finally, a spatial EPI encoding is employed to acquire the images. Three gradients are implemented in the three spatial directions for slide selection $G_{SS}$, read orientation $G_{RO}$ and phase encoding $G_{PE}$. The NOGSE gradients can be used in any orientation, although we applied them in the read orientation direction.
		\label{Figure_3}}
\end{figure*}

MRI scans were performed at 9.4T on a Bruker Avance III HD WB NMR scanner with a 1H resonance frequency of $\omega_z$\,=\,400.15\,MHz. We used a Bruker Micro 2.5 probe capable of producing gradients of up to 1500\,mT/m in the three spatial directions. We programmed and implemented with ParaVision 6 the NOGSE MRI sequences shown in Figure~\ref{Figure_3}~\cite{capiglioni_noninvasive_2021}. All experiments have the following sequence parameters: the Repetition time TR\,=\,2000\,ms; Field of view (FOV)\,=\,16$\times$16\,mm$^2$, with a matrix size of 192$\times$192, leading to an in-plane resolution of 83$\times$83\,$\mu$m$^2$; and slice thickness of 1\,mm.

For experiments involving the analysis of the NOGSE signal \textit{vs} the duration $t_C$ in Sections~\ref{Sect_D0_estimation} and \ref{Sect_SizeDistr_estimation}, the images were acquired using echo planar imaging (EPI) encoding with 4 segments with a scan time of $\sim$\,4\,minutes with 16 averages and an echo time TE\,=\,70.5\,ms; a diffusion time $t_D$ varying from 12.5\,ms to 24.5\,ms, although in most experiments $t_D$ was fixed at 21.5\,ms. We used $2<N<20$ for sharp modulations and $4<N<20$ for smooth modulations. The time $t_C$ for the lobes duration varied between 0.3\,ms and $t_D/N$; the gradient strength $G$ varied from 20\,mT/m to 600\,mT/m, with ramp times of 50\,mT/m/us, and were applied perpendicular to the main axis of the aramid fibre bundles.

For experiments involving the analysis of the NOGSE signal \textit{vs} the diffusion time $t_D$ (Section~\ref{Sect_shift_measurement}), the images were acquired with an EPI encoding with 4 segments, with an image scan time of $\sim$\,2.5 hours with 512 averages and an echo time TE\,=\,171.5\,ms; a diffusion time $t_D$ varied from 1.5\,ms to 80\,ms. We used $N$\,=\,4 for sharp and smooth modulations. The lobe duration $t_C$\,=\,0.3\,ms was the shortest time used to approximate the Hahn-limit modulation for NOGSE and $t_C$\,=\,$t_D/N$ for the CPMG-limit NOGSE modulation. The gradient strength $G$ was $G_{sharp}$\,=\,100\,mT/m and $G_{smooth}$\,=\,141\,mT/m for the sharp and smooth gradient modulations, respectively, with ramp times of 50\,mT/m/us. These values were chosen to make the signal decay rate in the restricted diffusion regime equal (see Equations~\ref{Eq_BetaRestrHahnC} and \ref{Eq_BetaRestrHahnS}). The gradients were applied perpendicular to the main axis of the aramid fibre bundles.

We determined the relaxation times $T_1$\,=\,1794.5\,ms, $T_2$\,=\,70.41\,ms, and $T_2^*$\,=\,16.294\,ms in ROI1 using RARE-VTR, MSME and MGE sequences, respectively. The free water relaxation times were determined from ROI5, giving $T_1$\,=\,3409.4\,ms, $T_2$\,=\,154.4\,ms, $T_2^*$\,=\,55.181\,ms.

The NOGSE signal arising from each voxel and ROI of the acquired images was analysed and plotted as a function of the NOGSE's acquisition parameters $t_C$, $t_D$, $N$ or $G$. Fittings to the experimental data were done based on the NOGSE theoretical model considering Equation~\ref{Eq_MagnDistr} with the microstructure size parameters that describe the distribution, \textit{i.e.} the mean size $l_c$ and standard deviation $\sigma_c$. A uniform free diffusion coefficient $D_0$\,=\,2.30\,$\mu$m$^2$/ms of water molecules at room temperature 25\,°C was used based on its experimental estimation using the signal arising from ROI5 (see Section~\ref{Sect_D0_estimation}).

\subsection{NOGSE sensitivity to microstructure-size estimates}
\label{Sect_NOGSE_sensitivity}
To evaluate the optimal acquisition setup for estimating microstructure restriction-size distributions, we use an information gain metric $S$~\cite{jimenez_optimization_2023}. It is the derivative of the mean square distance of the NOGSE curve with respect to changes in the parameters to be estimated, as these are obtained by fitting a theoretical representation to the experimental data based on reducing the mean square distance between the corresponding points. The highest derivative predicts the optimal NOGSE acquisition parameters for a given size distribution, consistently with quantum information estimation bounds~\cite{zwick_precision_2020,jimenez_optimization_2023}.

More precisely, the measured magnetisation signal $M(\theta,\epsilon)$ is defined by the acquisition parameters $\theta=\{t_d,G,N\}$ and the parameters that characterise the underlying lognormal restricton-size distribution of the sample under study, $\epsilon=\{\mu,\sigma\}$ (see Equation~\ref{Eq_LogNormal}). Thus, the aim is to find an optimal set of parameters $\theta$ to measure $\epsilon$ with the greatest precision possible. 

We can define the information gain metric $S$ as the sensitivity of the result of a measurement against perturbations $\epsilon'=\{\mu',\sigma'\}$ in the real parameters to be measured $\epsilon=\{\mu,\sigma\}$.

The optimal acquisition parameters $\theta=\{t_d,G,N\}$ are those that maximise the $D_{L2}(\theta,\epsilon,\epsilon')$ distance for small variations $\epsilon'=\{\mu',\sigma'\}$ in the vicinity of the true values $\epsilon=\{\mu,\sigma\}$. This $D_{L2}$ distance is the canonical distance for Lebesgue square-integrable functions (L2), since we consider the signal measurement as a function of the gradient-lobe duration $t_c$, and it is given by
\begin{equation}
	D_{L2}(\theta,\epsilon,\epsilon')=\sqrt{ \int_{Hahn}^{CPMG} [M(\theta,\epsilon,t_C) - M(\theta,\epsilon',t_C)]^2 dt_C }.
	\label{Eq_Distancia}
\end{equation}

As we aim to optimise two acquisition parameters -- the diffusion time $t_D$ and gradient strength $G$ -- we define the information gain metric 
\begin{equation}
	S(\theta,\epsilon,\Delta):=\overline{D_{L2}}\big\rvert_{C(\epsilon,\Delta)},
	\label{Eq_Svty}
\end{equation}
as the mean of the distance $D_{L2}$ over an ellipsoidal curve $C(\epsilon,\Delta)$, which is given by
\begin{equation}
	C(\epsilon,\Delta) := \epsilon\left[ 1+ \Delta \left( \cos(k),\sin(k) \right) \right], \, 0\leq k \leq 2\pi.
	\label{Eq_Curva}
\end{equation}
This curve $C(\epsilon,\Delta)$ parametrises an ellipse with minor and major radius $r_1=\mu \Delta$ and $r_2=\sigma \Delta$. The parameter $\Delta$ defines the $\epsilon'$ over which the $D_{L2}$ distance will be calculated. The minimum value $\Delta=0$ corresponds to taking $\epsilon'=\epsilon$ and the maximum value $\Delta=1$, to $\epsilon'=0$ and $\epsilon'=2\epsilon$.

In this way, we obtain the sensitivity of the result of a measurement to changes in the set of parameters as a function of $\theta$. Therefore, the optimal control parameters will be those that maximise this information gain metric $S$.

$T_2$ relaxation was included in the calculation of the information gain metric $S$ assuming ideal modulations of the NOGSE sequences and echo time $TE$, without accounting for the noise introduced on the signal acquisition process. The echo time was calculated as $TE$\,=\,$2t_D$\,+\,$t_p$, with $t_p$ being the time left for the RF pulses and crusher gradients. The maximum diffusion time $t_D$ considered was $T_2/2$. 

\section{Results and Discussion}
\label{Results} 
\subsection{Optimising Acquisition Parameters for Accurate Estimation of Restriction-Size Distributions}
\label{Sect_optimal_sensitivity}
We first performed a preliminary estimation of the restriction size distribution in ROI1 of the fibre phantom (Figure~\ref{Figure_2}c) to determine the NOGSE optimal configuration using the method described in Section~\ref{Sect_NOGSE_sensitivity}. We obtained $l_c$\,=\,7.3\,$\mu$m and $\sigma_c$\,=\,2.8\,$\mu$m from a fitting of the theoretical model of Equation~\ref{Eq_MagnDistr} to the experimental data. These values of $l_c$ and $\sigma_c$ are the averaged parameters that describe the size distributions obtained using the smooth NOGSE modulation, with $N$\,=\,16, and several measasurements performed within the diffusuon time $t_D$ and gradinet strength $G$ ranges 40\,-\,60\,ms and 70\,-\,300\,mT/m, respectively. They were used as reference values to predict the optimal values of $N$, $G$ and the NOGSE diffusion time $t_D$ for each type of gradient modulation using the protocol described in Section~\ref{Sect_NOGSE_sensitivity}.

According to previous works, the number of refocusing periods should be kept to a minimum for a better inference of the microstructure-size restriction parameter~\cite{zwick_precision_2020}. We found consistent results in the numerical evaluation; therefore, we kept the number of refocusing periods at the lowest achievable value. According to the available NOGSE modulation sequence design (Figure~\ref{Figure_1}), the lowest $N$ used are 2 and 4 for the sharp and smooth gradient modulation, respectively.

\begin{figure*}[ht]
	\centerline{\includegraphics[width=\textwidth]{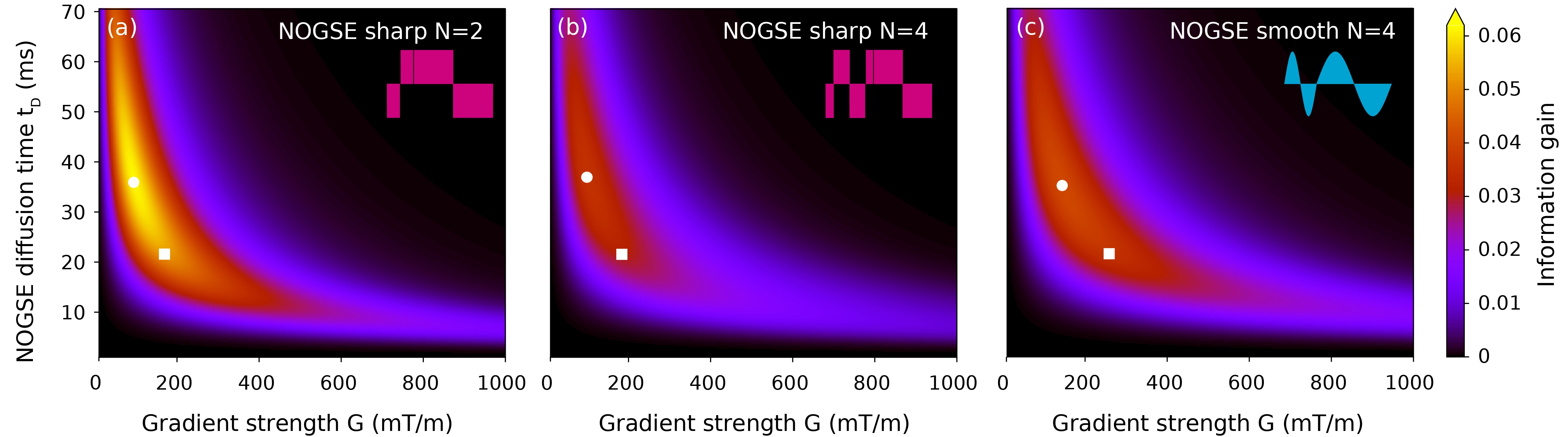}}
	\caption{Information gain maps obtained for sharp and smooth NOGSE modulations that predict the optimal acquisition parameters for each gradient modulation sequence, given by the highest gain in each case to a given restriction-size distribution. To generate these maps we obtained a preliminary estimate of the restriction-size distribution in ROI1 of the fibre phantom, given by a mean size $l_c$\,=\,7.3\,$\mu$m and standard deviation $\sigma_c$\,=\,2.8\,$\mu$m. We considered $T_2$\,=\,70.41\,ms, $D_0$\,=\,2.30\,$\mu$m$^2$/ms and $TE$\,=\,2$t_D$\,+\,27.5\,ms. The white circles show the optimal $t_D$ and $G$ obtained for the entire range of acquisition parameters covered to generate each map. The optimal gradient strength at $t_D$\,=\,21.5\,ms is marked with white squares in the three maps.
		\label{Figure_4}}
\end{figure*}

Then, we performed a systematic exploration of the information gain metric described in Section~\ref{Sect_NOGSE_sensitivity}, as a function of $G$ and $t_D$. Figure~\ref{Figure_4} shows three information gain maps for the NOGSE modulations with $N_{sharp}$\,=\,2 or $N_{sharp}$\,=\,4 and $N_{smooth}$\,=\,4, respectively. They show the optimal values of the acquisition parameters $t_D$ and $G$ to estimate the restriction size distribution, corresponding to the highest information gain. For the NOGSE sharp modulation: $t_{D,\,opt}$\,=\,35.9\,ms and $G_{opt}$\,=\,94.6\,mT/m for $N$\,=\,2, and $t_{D,\,opt}$\,=\,37\,ms and $G_{opt}$\,=\,99.5\,mT/m for $N$\,=\,4. Whereas for the NOGSE smooth modulation: $t_{D,\,opt}$\,=\,35.2\,ms and $G_{opt}$\,=\,144.3\,mT/m. Although we find the optimal diffusion time $t_D$ around 36\,ms, we used a lower value to improve signal-to-noise ratio (SNR) in all images. By experimental observation, we found $t_D$\,=\,21.5\,ms to have a good compromise between the predicted optimal value and the SNR of the images.

At $t_D$\,=\,21.5\,ms, the predicted optimal gradient strengths for the sharp modulations are $G_{sharp}$\,=\,169\,mT/m with $N_{sharp}$\,=\,2 and $G_{sharp}$\,=\,184\,mT/m if $N_{sharp}$\,=\,4, and for the smooth modulation $G_{smooth}$\,=\,259\,mT/m with $N_{smooth}$\,=\,4. As evidenced in Figure~\ref{Figure_4}, there is a \textit{plateau} on the information gain where NOGSE reaches a comparable information gain (yellow-red region). Hence, we expect a similar estimation accuracy of the underlying microstructure size distribution by choosing $t_D$ and $G$ within this region. Furthermore, we do not find a significant difference between using sharp or smooth modulations. However, the required gradient strength increases for the smooth modulations, and gives a slightly better precision.

\subsection{Estimation of the free diffusion coefficient}
\label{Sect_D0_estimation}
\begin{figure*}[ht]
	\centerline{\includegraphics[width=\textwidth]{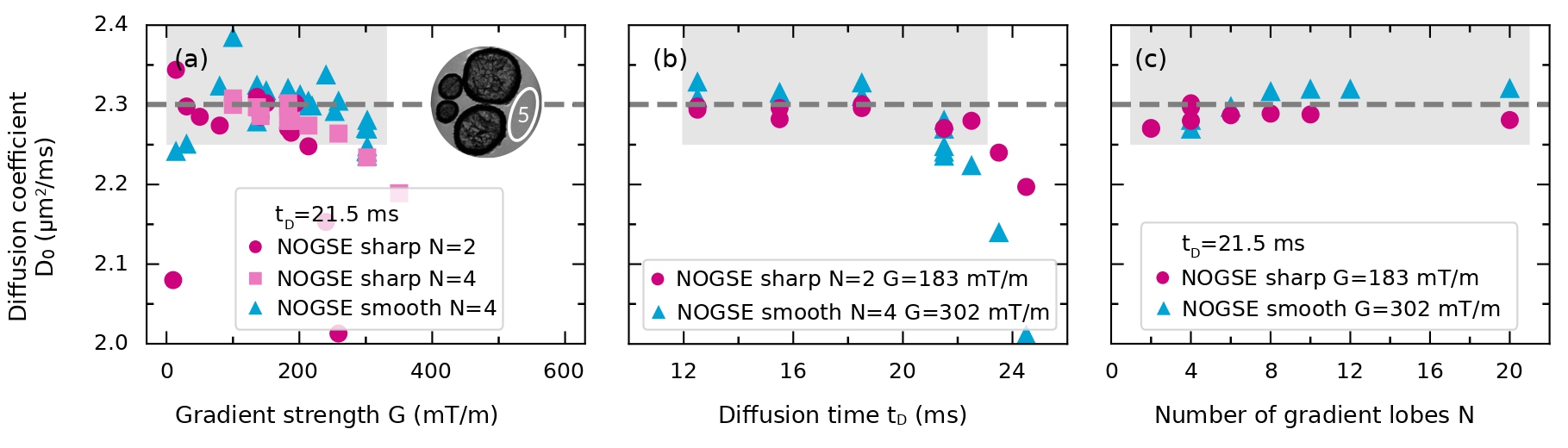}}
	\caption{Estimated free diffusion coefficient $D_0$ of free water in ROI5 of the phantom (inset) described in Figure~\ref{Figure_2}. The $D_0$ as a function of (a) the gradient strength $G$, (b) the NOGSE diffusion time $t_D$ and (c)  the total number of gradient lobes $N$. The grey dashed line in each plot at $D_0$\,=\,2.30\,$\mu$m is the average of all the experiments in the optimal regions, \textit{i.e.} determined by the dynamic range with fewer fluctuations in the estimated value of the diffusion coefficient (shown with shaded grey boxes), in line with the literature.
		\label{Figure_5}}
\end{figure*}

We estimate the free diffusion coefficient $D_0$ based on the free water signal in ROI5 (Figure~\ref{Figure_2}c). When the diffusion time is much shorter than $\tau_c$, we achieve the free diffusion limit. The NOGSE magnetisation decays with the attenuation factor
\begin{multline}
	\beta^{free}_{NOGSE \, sharp}=\\
	\frac{\gamma^2D_0G^2}{12}\left( \left( N-1 \right)t_C^3 + \left( t_D- \left( N-1 \right)t_C \right)^3 \right),
	\label{Eq_NOGSEcuad_libre}  
\end{multline}
for the sharp gradient modulation and
\begin{multline}
	\beta^{free}_{NOGSE \, smooth}=\\
	\frac{3\gamma^2D_0G^2}{8\pi^2} \left( 4\left( N-2  \right)t_C^3 + \left( t_D-\left( N-2 \right)t_C \right)^3 \right),
	\label{Eq_NOGSEsin_libre}
\end{multline}
for the smooth NOGSE modulation. Notice that both expressions are independent of $\tau_c$, and consequently, of the restriction size $l_c$. We use these expressions to estimate $D_0$. Figure~\ref{Figure_5} shows the estimated $D_0$ values as a function of the NOGSE acquisition parameters $G$, $t_D$ and $N$, for the smooth and sharp modulations.

Figure~\ref{Figure_5} demonstrates that there is a range for the values of the acquisition parameters $G$, $t_D$ and $N$ where the estimated $D_0$ remains stable around $D_0$\,=\,(2.30\,$\pm$\,0.03)\,$\mu$m$^2$/ms (averaged value of all the experiments performed in the shaded regions of the figure). These regions are thus more sensitive for accurately estimating $D_0$.

Outside of these regions, for gradient strengths $G$ exceeding approximately 300\,mT/m, the signal experiences low SNR, resulting in an inefficient estimation. Similarly, excessively long diffusion times reduce SNR, leading to a diminished estimation efficiency. In contrast, for very low gradient strengths, selective RF pulses and image encoding introduce diffusion weighting effects, eclipsing the NOGSE diffusion weighting. The variation of the number of gradient lobes within the explored dynamic range does not yield significant changes. Nonetheless, a slight improvement is observed by increasing $N$.

We use the estimated free diffusion coefficient $D_0$\,=\,2.30\,$\mu$m$^2$/ms to obtain the restriction size distributions present in the different aramid fibre bundles of the phantom. Notice that the optimal acquisition parameters to estimate restriction size distributions are not in general the same as those identified for estimating $D_0$. For instance, an optimal diffusion time $t_D$ to sense restriction size distributions needs to be long enough to allow water molecules to fully explore the compartment restrictions~\cite{zwick_precision_2020}. On the contrary, the optimal inference of the free diffusion coefficient does not require the water molecules to sense the compartment confinement.

\subsection{Validating NOGSE decay-shift predictions: a proof of concept}
\label{Sect_shift_measurement}
\begin{figure*}[ht]
	\centerline{\includegraphics[width=\textwidth]{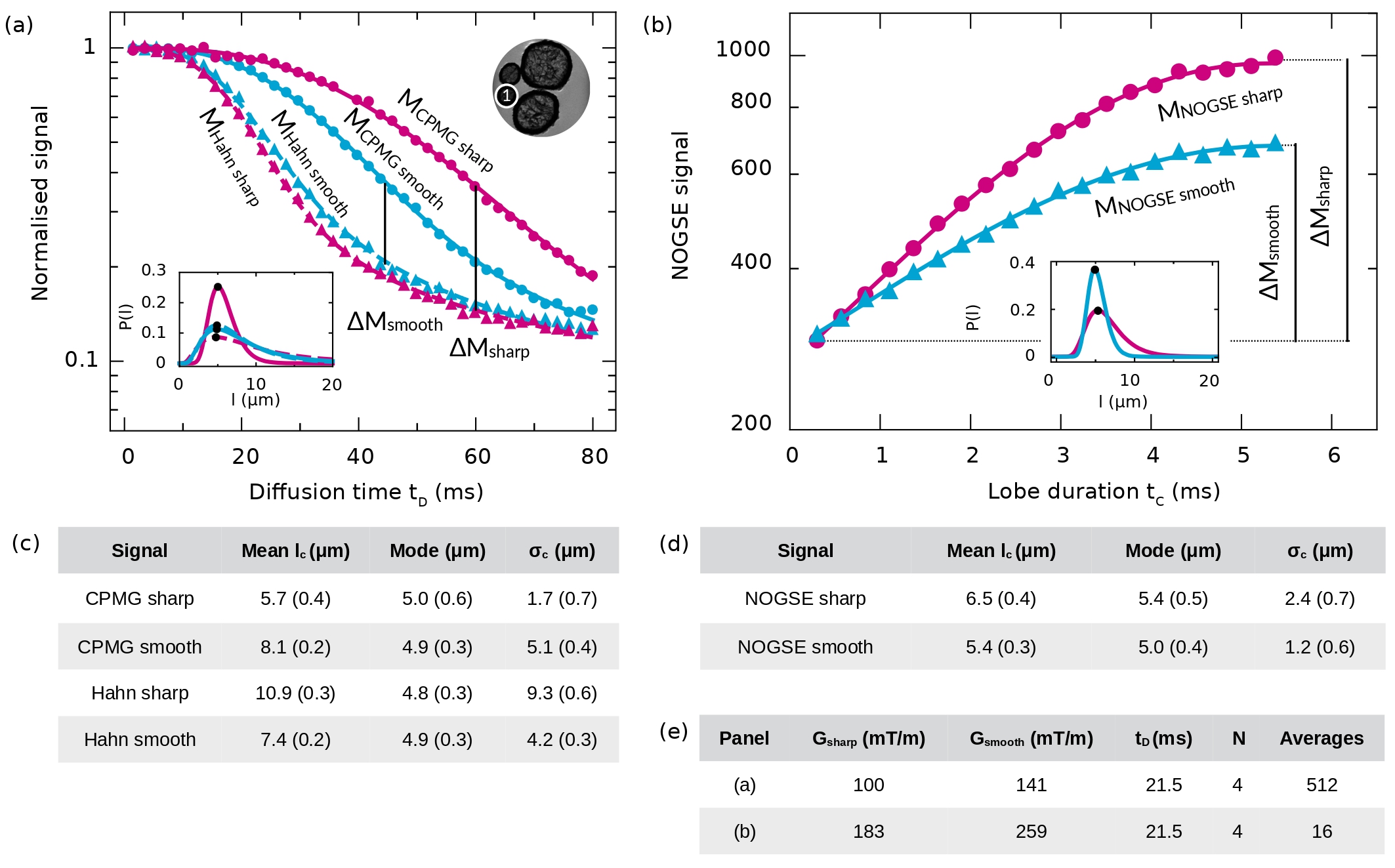}}
	\caption{(a)~Signal decay as a function of the diffusion time $t_D$ at the limiting values of the lobe duration $t_C$\,=\,0.3\,ms (Hahn modulation, triangles) and $t_C$\,=\,$t_D/N$ (CPMG modulations, circles) and~(b)~NOGSE signal as a function of the lobe duration $t_C$, for sharp (pink) and smooth (light blue) gradient modulations. For both figures, the experimental data is extracted from ROI1 of the fibre phantom (top-right inset from panel (a)). The dashed and solid lines are fitted curves to the experimental data considering a lognormal size distribution in the model and $D_0$\,=\,2.30\,$\mu$m$^2$/ms. The bottom insets show the reconstructed lognormal restriction-size distributions with the same line styles as in the main panels. In the inset from panel (a), indistinguishable distribution-size probabilities are given by the dashed and solid, light-blue curves. (c, d) Estimated restriction-size distributions parameters from (a) and (b), respectively. (e) NOGSE setup parameters.
		\label{Figure_6}}
\end{figure*}

We experimentally assess the manifestation of the NOGSE decay-shift by applying both gradient modulations on the white-matter phantom. Figure~\ref{Figure_6}a displays the experimental measurements of the NOGSE signal decay, depicted in pink for the sharp modulation and light blue for the smooth modulation. The NOGSE decay is shown for the limiting values of $t_C\rightarrow0$ (Hahn modulation, triangles) and $t_C$\,=\,$t_D/N$ (CPMG modulation, circles) with $N$\,=\,4 as a function of the total diffusion time $t_D$, for both sharp and smooth gradient modulations.

The selected acquisition parameters for the NOGSE sequence setup ensure that both modulations yield the same decay rate within the restricted diffusion regime for a single restriction size. Based on Equations~\ref{Eq_DeltaMrestr_cuad} and \ref{Eq_DeltaMrestr_sin}, this condition is achieved when $G_{smooth}$\,=\,$\sqrt{2}G_{sharp}$. Hence, we chose $G_{sharp}$\,=\,100\,mT/m and $G_{smooth}$\,=\,141\,mT/m as in Figures~\ref{Figure_1}b and d.

Comparing the signal decays shown in Figure~\ref{Figure_6}a with the ones predicted by the theoretical model for a single restriction size in Figures~\ref{Figure_1}b and d, we find that the overall behaviour of the predicted shifts is observed. The main difference is due to the restriction size distribution effects on the experimental data regardless of the chosen modulation type. Each restriction size from the probability distribution modelled by Equation~\ref{Eq_MagnDistr}, contributes with a different decay rate and shift, which are manifested differently as a function of the diffusion time $t_D$. Still, in the region of the normalised signal within 0.3 and 0.6 an effective restriction size seems to dominate on the decaying curve manifesting the expected shift of the exponentially decaying curves as in Figures~\ref{Figure_1}b and d. At longer times, the spin signal contribution from smaller restrictions becomes dominant and the decaying laws change. Moreover, Rician noise effects may also become significant.

The NOGSE signal shift reduces as a function of the diffusion time $t_D$ significantly faster for the smooth modulation compared with the sharp modulation as expected. For the sharp modulation, it also seems to be reduced. In this latter case, it might be due to the fact that the sharp modulations are performed with trapezoidal pulse gradients, which are not instantaneous as in the theoretical model. It might also be due to the size-distribution effects or Rician noise manifestation.

The fit of the theoretical representations to the experimental data provides comparable size distributions for both types of gradient modulations. The modes of the distributions are shown with black dots which are indistinguishable between the different cases. However, the distribution width evidences some variations (see the caption of Figure~\ref{Figure_6}). We provide further discussions about these differences in the following section.

We also present the NOGSE signal as a function of the lobe duration $t_C$ in Figure~\ref{Figure_6}b. As predicted, the NOGSE contrast $\Delta M$ is larger for the sharp modulation. As shown in the inset, the extracted size distributions from the fitting of the theoretical model of Equation~\ref{Eq_MagnDistr} to the experimental data are also comparable. Again, the mode is robust, \textit{i.e.} they are indistinguishable in both cases, yet there are differences on the distribution widths.

\subsection{Estimation of restriction-size distributions}
\label{Sect_SizeDistr_estimation}
\begin{figure*}[ht]
	\centerline{\includegraphics[width=\textwidth]{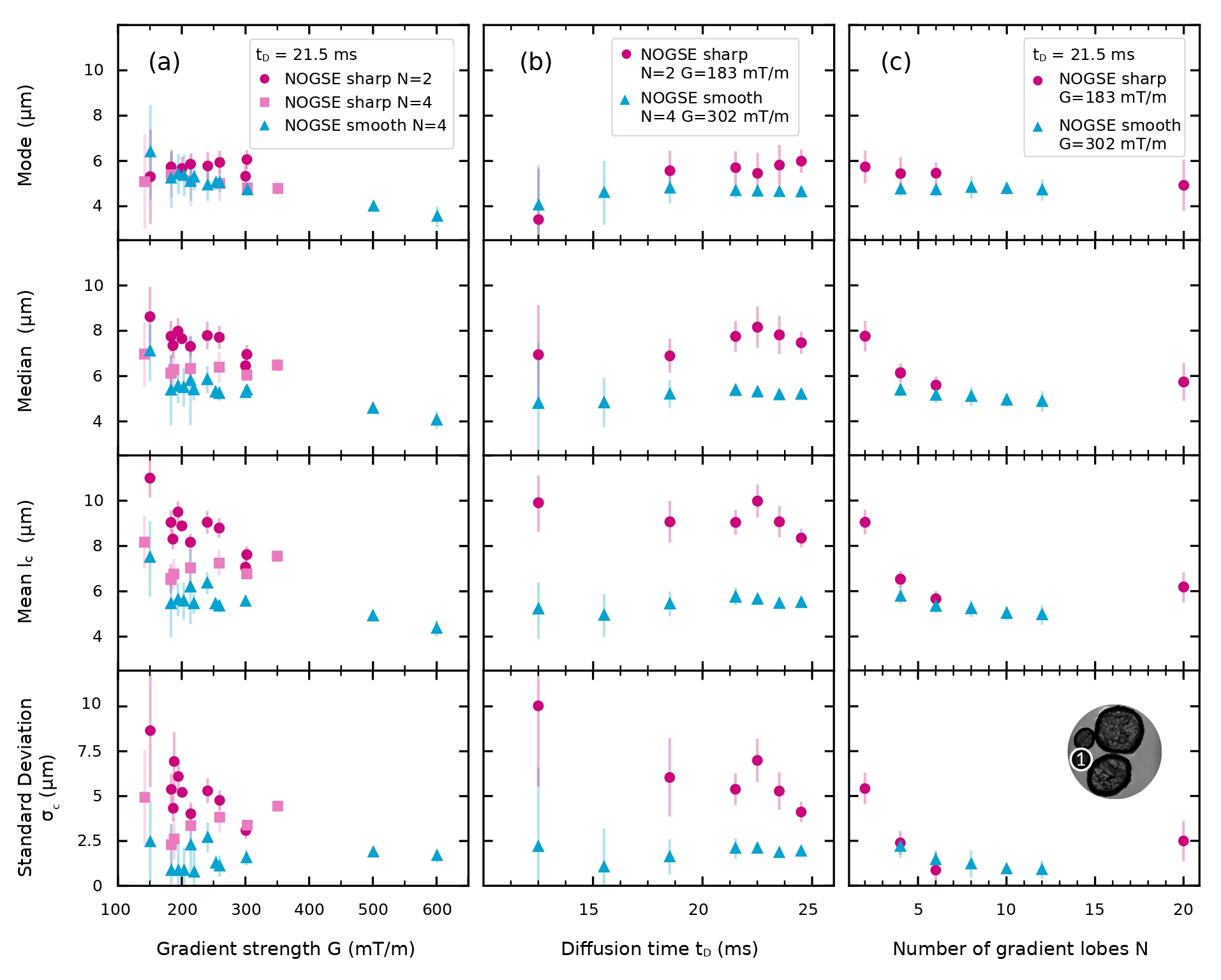}}
	\caption{Extracted parameters that characterise the restriction size distributions for different NOGSE sequences. The experimental data obtained from ROI1 (lower-right inset) was fitted with a NOGSE signal time evolution assuming a lognormal distribution. The mean size $l_c$ and standard deviation $\sigma_c$ were fitted and, then, the mode and median were estimated. Mode, median, mean, and standard deviations are shown as a function of (a) the gradient strength $G$, (b) the NOGSE diffusion time $t_D$ and (c) the number of gradient lobes $N$ at the fixed setup parameters indicated at the top of each panel.
		\label{Figure_7}}
\end{figure*}

We performed a more systematic evaluation of the reconstructed restriction size distributions determined from the different NOGSE sequences using sharp and smooth gradient modulations. We carried out NOGSE experiments for different $t_D$, $N$ and $G$ as a function of the lobe duration $t_C$, as shown in Figure~\ref{Figure_6}b. Figure~\ref{Figure_7} shows the extracted parameters that characterise the corresponding estimated size distributions for all the considered cases. We present the mode, median, mean size and standard deviation $\sigma_c$ of the restriction size distribution obtained from the fitted curves to the experimental data of the averaged signal of ROI1. The symbols represent the fitted size distribution parameters as a function of different NOGSE modulations and acquisition parameters: $G$ (Figure~\ref{Figure_7}a), $t_D$ (Figure~\ref{Figure_7}b) and $N$ (Figure~\ref{Figure_7}c).

For both modulation types, the mean size and standard deviation have more dispersion than the mode and, in second place, than the median. The mode evidences to be a more robust parameter for characterising the restriction-size compared to the mean size. For all the cases, including variations of $G$, $t_D$ and $N$, the mode is quite stable near to the optimal setup parameters of the sequence for the restriction-size inference predicted in Section~\ref{Sect_optimal_sensitivity}. It is even robust when comparing the different modulation shapes of the gradient. However, the mean and median sizes do not manifest robustness, as they are dependent on the value of the estimated standard deviation $\sigma_c$.

As the sequence is not sensitive to sizes larger than those explored during the diffusion time, the estimation of size distributions appears to incline towards a distribution that encompasses larger sizes beyond the range of exploration due to the lack of sensitivity to them. The estimation of larger sizes is correlated with the inclusion of smaller sizes when increasing $\sigma_c$. For this reason, despite the dispersion of the distribution shape, the mode seems always to be a robust parameter to characterise the system.

Both NOGSE modulations estimate similar lognormal distributions, although we find that, in general, smooth gradient modulations give slightly smaller microstructure-size mode and mean values, as well as for the distribution widths. Similar qualitative behaviours are observed for the extracted size distribution parameter from different ROIs in the phantom.

The estimated values have less dispersion when estimated with NOGSE acquisition parameters near to the optimal region predicted in Section~\ref{Sect_optimal_sensitivity}. Notice that higher parametric fluctuations are observed at the lower and higher ends of the gradient strengths and also at the extremes of the explored range for the diffusion time. These extremes are less sensitive for estimating the size distribution parameters as they are far from the optimal setup for the inference. We find, in general, that the regions with less dispersion in the estimated size distribution parameters, are at a bit higher gradient strengths compared with the predicted optimal value. In particular this is more noticeable for the sharp modulation. This seems to be consistent with the fact that the numerical predictions assume ideal squared shapes, while in practice they have trapezoidal shapes. By increasing $N$, we find that the estimation provides slightly lower restriction sizes. This is consistent with the fact that the lower $N$ possible is the most sensitive for estimating restriction sizes~\cite{zwick_precision_2020} and that increasing $N$ for the same diffusion time reduces the maximum restriction sizes that can be estimated.

\begin{figure*}[ht]
	\centerline{\includegraphics[width=\textwidth]{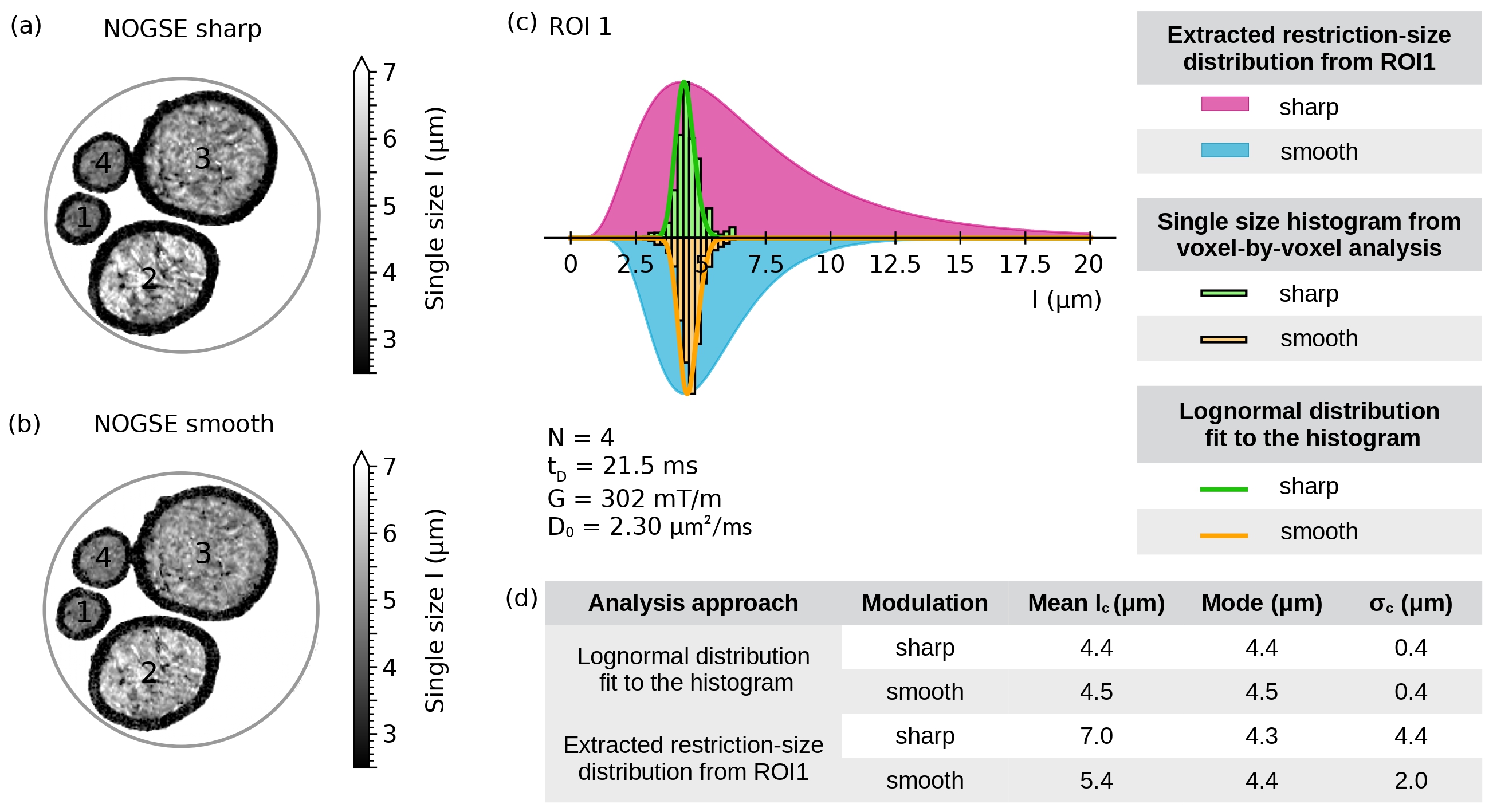}}
	\caption{Maps of single restriction sizes generated and analysed using fitted models based on images obtained with NOGSE in each voxel (83$\times$83$\times$1000\,$\mu m^3$), using (a) sharp and (b) smooth gradient modulations with the same setup parameters. (c) Histograms of the single restriction-size estimation in ROI1 from the maps in panels (a) and (b) are compared to the corresponding reconstructed lognormal restriction-size distribution extracted from fittings to the averaged signal of ROI1. (d) Lognormal distribution fit to the histograms from panel (c) and the extracted size-distribution parameters from the averaged signal fitting of ROI1.
		\label{Figure_8}}
\end{figure*}

In addition to evaluate the consistency of the estimations, we compare the resulting restriction size distributions obtained from the averaged signal of a ROI as shown in Figure~\ref{Figure_7} with the ones obtained from single-restriction size fits from the voxel-by-voxel data. In this case, we use the NOGSE parameters derived from the region in Figure~\ref{Figure_7}a that exhibits the lowest dispersion of the estimated values. Figure~\ref{Figure_8}a,b are maps of the determined single-restriction sizes for each voxel, for the sharp and smooth NOGSE modulations, respectively. The observed variations in fibre densities, based on the restriction size estimation, are consistent with the level of compaction achieved during the phantom preparation process. Notice that the assumption here is that the size-distribution width at every voxel is negligible.

Single-size histograms from the voxel-by-voxel fits of ROI1 are shown in Figure~\ref{Figure_8}c. Smooth and sharp modulations provide very similar size distributions. We also compare each histogram with the restriction-size distribution that arises from the averaged-signal fit of ROI1, for both types of gradient modulation. The distribution from the voxel-by-voxel differs from the one obtained via the averaged-signal. The distribution size mode is robust as both approaches give indistinguishable values. Nevertheless, there is an inconsistency in the distribution-width determination.

\section{Conclusions}
\label{Conclusions} 
Motivated by the need for improved diagnostic tools in neurodegenerative diseases, particularly in those associated with white-matter alterations, our study is focused on determining microstructural sizes using the NOGSE sequence in DWI. NOGSE has shown to be a tool to selectively probe microstructure sizes based on a decay-shift, especially using sharp gradient modulations. Our research aimed to evaluate the applicability of this technique focusing on comparing the estimation of microstructure size-distributions using both sharp and smooth gradient modulations, where the latter is more suitable for clinical protocols.

For this purpose we designed home-made phantoms to mimic axon bundles, in particular to study the diffusion within the extra-axonal compartments. We demonstrated with a combination of analytical derivations and numerical simulations using information theory tools, contrasted with proof-of-principle experiments, that both modulations can be used to estimate restriction-size distributions using NOGSE and its decay-shift. We found that both types of modulation give comparable microstructural information.

By a systematic analysis of the estimated microstructural properties of the white-matter phantom, we evaluate the reliability and self-consistency of obtaining compartment-size distributions with the NOGSE sequence. We found that the estimation of the distribution's mode is very robust, and thus reliable for both, sharp and smooth gradient modulations. On the contrary, the mean, median size and its distribution width are less robust. While further work needs to be performed, this seems to be associated with indistinguishable properties of the assumed log-normal distribution than to a specific method for the inference.

Overall, the presented results are a step towards improving diagnostic tools based on quantitative imaging of tissue microstructure that might allow early stage diagnosis of neurodegenerative diseases associated with axon alterations in the white matter of the brain.

\section*{Acknowledgements} \label{sec:acknowledgements}
This work was supported by CNEA; CONICET, ANPCyT-FONCyT PICT-2017-3156, PICT-2017-3699, PICT-2018-4333, PICT-2021-GRF-TI-00134, PICT-2021-I-A-00070; PIP-CONICET (11220170100486CO); UNCUYO SIIP Tipo I 2019-C028, 2022-C002, 2022-C030; Instituto Balseiro; Collaboration programs between the MINCyT (Argentina) and, MAECI (Italy) and MOST (Israel); and Erasmus+ Higher Education program from the European Commission between the CIMEC (University of Trento) and the Instituto Balseiro (Universidad Nacional de Cuyo).

\appendix*
\appendix
\section{NOGSE magnetisation decay\label{app1}}
As mentioned in the \textit{Theory} section of the main text (Subsection~\ref{DWMR}), the magnetisation signal observed from an ensemble of non-interacting and equivalent spins under the effects of a modulated gradient sequence is $M(t_D)$\,=\,$M(0)$\,$\langle$\,$e^{-i\phi(t_D)}\rangle$. Here, $\phi(t_D)$\,=\,$\int_{0}^{t_D}\omega(t)dt$ is the phase acquired by a spin during the evolution time $t_D$, and the brackets denote the average over the random phases of the spin ensemble~\cite{hahn_spin_1950,carr_effects_1954}. For the considered modulated gradient spin-echo sequences, the average phase becomes $\langle \phi(t) \rangle$\,=\,0. Then, as $\phi(t)$ typically follows a Gaussian distribution~\cite{klauder_spectral_1962,stepisnik_analysis_1981,stepisnik_validity_1999}, the signal will evidence a decay depending on the random phase variance as
\begin{equation}
	M(t_D)=M(0) e^{-\beta(t_D)} = M(0) e^{-\frac{1}{2}\langle \phi^2(t_D)\rangle}.
	\label{Eq_Magn_App}
\end{equation}
The attenuation factor $\beta(t_D)$ can be written in a Fourier representation as~\cite{stepisnik_time-dependent_1993,lasic_displacement_2006,shemesh_measuring_2013,alvarez_coherent_2013,zwick_maximizing_2016}
\begin{equation}
	\beta(t_D)=\frac{\gamma^2}{2}\int_{-\infty}^{\infty}\left|  F(\omega,t_D) \right|^2 S(\omega) d\omega.  
	\label{Eq_beta_App}
\end{equation}
Within this FT representation, the signal decay is determined by the integral of the overlap between the filter function $F(\omega,t_D)$ and the spectral density $S(\omega)$. The filter $F(\omega,t_D)$ only depends on the setup parameters of the chosen MGSE sequence, since it is defined by the FT of the gradient modulation $G(t)$. The spectral density $S(\omega)$ only depends on the characteristics of the diffusion-driven fluctuations, given by
\begin{equation}
	S(\omega)= \frac{D_0 \tau_c^2}{(1+\omega^2 \tau_c^2)\pi},
	\label{Eq_DensidadEspectral_Ap}
\end{equation}
where $D_0$ is the free diffusion coefficient and $\tau_c$ is the time required for most molecules to fully probe the compartment boundaries. Considering the Einstein’s expression $l_c^2$\,=\,$2 D_0 \tau_c$, the characteristic time $\tau_c$ is related to the restriction length $l_c$ of the microstructure compartment in which diffusion is taking place~\cite{callaghan_translational_2011}. Therefore, by suitably choosing the oscillating gradient modulation parameters to steer the NMR signal decay, one can probe the water diffusion process and, consequently, extract microstructural restriction sizes from its time dependence~\cite{gore_characterization_2010,shemesh_measuring_2013,alvarez_coherent_2013,capiglioni_noninvasive_2021}.

In the following Subsections we specify the signal decays expressions for the sharp and smooth gradient modulations used in this work (see Figures~\ref{Figure_1}a and c for a single diffusion restriction size $l_c$).

\subsection{NOGSE signal decay for sharp gradient modulations\label{app_cuad}}
For the ideal NOGSE sequence shown in Figure~\ref{Figure_1}a of the main text, gradient lobes of duration $t_C$ are repeated with alternated sign $N-1$ times giving the CPMG modulation and a last refocusing period which corresponds to the Hahn counterpart. The total diffusion duration for the CPMG part is $t'_C=(N-1)t_C$ and for the Hahn part, $t_H=t_D-(N-1)t_C$. For the sharp gradient modulation, the NOGSE magnetisation decay is~\cite{shemesh_measuring_2013,alvarez_coherent_2013}

\begin{multline}
	M_{NOGSE\,sharp}(t_D,t_{C},t_H,N,G) =\\ 
	M_{CPMG\,sharp}((N-1)t_C,N-1,G)   \\
	\times M_{Hahn\,sharp}(t_H,G)  
	\times M_{cross}(t_D,t_C,t_H,N,G),  
	\label{Eq_betacuad_Ap}
\end{multline}

with the Hahn signal decay $\beta_{Hahn\,sharp}(t_H,G)$ given by

\begin{multline}
	\beta_{Hahn\,sharp}(t_H,G)= \gamma^{2}G^{2}D_{0}\tau_{c}^{2}t_H\Bigg\{ \\ 
	1-\frac{\tau_{c}}{t_H}\left(3+\exp\left[-\frac{t_H}{\tau_{c}}\right]-4\exp\left[-\frac{t_H}{2\tau_{c}}\right]\right)\Bigg\},
	\label{Eq_betaHcuad_Ap}
\end{multline}

and for the CPMG signal decay $\beta_{CPMG\,sharp}(t'_C,N,G)$ by

\begin{equation}
	\beta_{CPMG\,sharp}(t'_C,N,G)=\gamma^{2}G^{2}D_{0}\tau_{c}^{2}\left(t'_C-\tau_{c}\left(A-B\right)\right),
	\label{Eq_betaCcuad_Ap}
\end{equation}

with
\begin{equation}
	A=\left(2N+1\right)-\left(-1\right)^{N}\exp\left[-\frac{t'_C}{\tau_{c}}\right]
	\label{Eq_betaCcuadA_Ap}
\end{equation}
and

\begin{multline}
	B=\frac{4}{\left(\exp\left[-\frac{t'_C}{N\tau_{c}}\right]+1\right)^{2}}    \Bigg\{ 
	(-1)^{N+1}\exp\left[-\frac{t'_C}{\tau_{c}}\right]\Bigg( \\
	\exp\left[-\frac{3t'_C}{2N\tau_{c}}\right]+\exp\left[-\frac{t'_C}{2N\tau_{c}}\right]-\exp\left[-\frac{t'_C}{N\tau_{c}}\right]\Bigg)  \\
	+ \exp\left[-\frac{3t'_C}{2N\tau_{c}}\right]+
	\exp\left[-\frac{t'_C}{2N\tau_{c}}\right]+
	\exp\left[-\frac{2t'_C}{N\tau_{c}}\right] \\ +
	\exp\left[-\frac{t'_C}{N\tau_{c}}\right](N+1)\Bigg\}.
	\label{Eq_betaCcuadB_Ap}
\end{multline}

The cross term $M_{cross}(t_D,t_C,t_H,N,G)$ has an argument

\begin{multline}
	\beta_{NOGSE\,sharp,\, cross}(t_D,t_C,t_H,N,G)=\frac{\gamma^{2}G^{2}D_{0}\tau_{c}^{3}}{\exp\left[\frac{t_C}{\tau_{c}}\right]+1}  \\
	\Bigg\{ 
	\Bigg(  1+\exp\left[-\frac{t_H}{\tau_{c}}\right]-2\exp\left[-\frac{t_H}{2\tau_{c}}\right]-
	2\exp\left[\frac{2t_C-t_H}{2\tau_{c}}\right]\\
	+\exp\left[\frac{t_C-t_H}{\tau_{c}}\right]   
	+4\exp\left[\frac{t_C-t_H}{2\tau_{c}}\right]-2\exp\left[\frac{t_C-2t_H}{2\tau_{c}}\right]\\  
	-2\exp\left[\frac{t_C}{2\tau_{c}}\right]+\exp\left[\frac{t_C}{\tau_{c}}\right]\Bigg)  \\
	+(-1)^{N}\Bigg( 
	\exp\left[-\frac{t_CN-t_C+t_H}{\tau_{c}}\right] \\
	-2\exp\left[-\frac{2t_CN-2t_C+t_H}{2\tau_{c}}\right]   
	-2\exp\left[-\frac{2t_CN-4t_C+t_H}{2\tau_{c}}\right]  \\
	+\exp\left[-\frac{t_CN-2t_C+t_H}{\tau_{c}}\right]  
	+4\exp\left[-\frac{2t_CN-3t_C+t_H}{2\tau_{c}}\right] \\
	-2\exp\left[-\frac{2t_CN-3t_C+2t_H}{2\tau_{c}}\right]   
	+\exp\left[-\frac{(N-1)t_C}{\tau_{c}}\right] \\
	-2\exp\left[-\frac{(2N-3)t_C}{2\tau_{c}}\right]+\exp\left[-\frac{(N-2)t_C}{\tau_{c}}\right]\Bigg)
	\Bigg\}.
	\label{Eq_betaCcuad_cross}
\end{multline}

The difference between the two limiting values, $t_C\rightarrow0$ and $t_C=t_D/N$ of this expression is the NOGSE sharp modulation contrast $\Delta M_{sharp}$.

When the diffusion times $t_C$ and $t_D - (N-1)t_C$ are much longer than $\tau_c$, the NOGSE magnetisation decay in the restricted diffusion regime is
\begin{multline}
	\beta_{NOGSE \,sharp}^{restr}(t_D,t_C,t_H,N,G)=\\
	\gamma^2D_0G^2\tau_c^2\Bigg(  t_D-\left( 1+2N \right)\tau_c   \Bigg)
	\label{Eq_betarestrcuad_Ap}
\end{multline}
from which we obtain the restricted diffusion expressions for the Hahn modulation at the limiting value $t_C\rightarrow0$ and for the CPMG modulation at $t_C$\,=\,$t_D/N$ (Equations~\ref{Eq_BetaRestrHahnC} and \ref{Eq_BetaRestrCPMGC}, respectively). With these expressions we obtain the NOGSE contrast $\Delta M_{sharp}^{restr}$ from Equation~\ref{Eq_DeltaMrestr_cuad}.

In the free diffusion regime, where the diffusion times are much shorter than $\tau_c$, the NOGSE magnetisation decays with
\begin{multline}
	\beta^{free}_{NOGSE \, sharp}(t_D,t_C,N,G)=\\ 
	\frac{\gamma^2D_0G^2}{12}\left( \left( N-1 \right)t_C^3 + \left( t_D- \left( N-1 \right)t_C \right)^3 \right).
	\label{Eq_NOGSEcuad_libre_AP}
\end{multline}

\subsection{NOGSE signal decay for smooth gradient modulations\label{app_SIN}}
Instead, if we consider a sinusoidal waveform as a representation of a smooth gradient modulation as in Figure~\ref{Figure_1}c, the NOGSE sharp magnetisation decay for a given restriction size $l_c$ results
\begin{multline}
	\beta_{NOGSE\,smooth}(t_D,t_{C},t_H,N,G) =\frac{D_{0} G^{2} \gamma^{2} \tau_{c}^{2}}{2} \Bigg\{ \\
	\frac{(N-2)t_{C}^{3}}{t_{C}^{2}+\pi^{2}\tau_{c}^{2}}
	+\frac{2\pi^{2}\tau_{c}^{3}t_{C}^{2}}{\left(t_{C}^{2}+\pi^{2}\tau_{c}^{2}\right)^{2}}\left(1-\exp\left[-\frac{(N-2)t_{C}}{\tau_{c}}\right]\right)  \\ 
	+\frac{t_H^{3}}{t_H^{2}+4\pi^{2}\tau_{c}^{2}}+\frac{8\pi^{2}\tau_{c}^{3}t_H^{2}}{\left(t_H^{2}+4\pi^{2}\tau_{c}^{2}\right)^{2}}\left(1-\exp\left[-\frac{t_H}{\tau_{c}}\right]\right)   \\
	-\frac{4\pi^{2}\tau_{c}^{3}xy}{\left(t_{C}^{2}+\pi^{2}\tau_{c}^{2}\right)\left(t_H^{2}+4\pi^{2}\tau_{c}^{2}\right)} \Bigg( 1+\exp\left[-\frac{t_{D}}{\tau_{c}}\right] \\
	-\exp\left[-\frac{(N-2)t_{C}}{\tau_{c}}\right] 
	-\exp\left[-\frac{t_H}{\tau_{c}}\right] \Bigg) \Bigg\},
	\label{Eq_betasin_Ap}
\end{multline}
The Hahn modulation for the smooth gradient modulation has a duration of $2t_H$\,=\,$t_D-(N-2)t_C$. The difference between the two limiting values, $t_C\rightarrow0$ and $t_C$\,=\,$t_D/N$ from this expression is the NOGSE smooth modulation contrast $\Delta M_{smooth}$.

When considering the restricted diffusion regime, the NOGSE magnetisation decay in this case results
\begin{multline}
	\beta_{NOGSE \,smooth}^{restr}(t_D,t_C,t_H,N,G)=\\
	\frac{D_{0}\gamma^{2}G^{2}\tau_{c}^{2}t_D}{2}\left(1-\frac{\left(N-2\right)\pi^{2}\tau_{c}^{2}}{t_D t_C}-\frac{4\pi^{2}\tau_{c}^{2}}{2t_D t_H}\right)
	\label{Eq_betarestrsin_Ap}
\end{multline}
from which we obtain the restricted diffusion expressions for the Hahn and CPMG extremes of Equations~\ref{Eq_BetaRestrHahnS} and \ref{Eq_BetaRestrCPMGS}, respectively, yelding the NOGSE contrast $\Delta M_{smooth}^{restr}$ from Equation~\ref{Eq_DeltaMrestr_sin}.

In the free diffusion regime, the NOGSE smooth magnetisation decays with
\begin{multline}
	\beta^{free}_{NOGSE \, smooth}(t_D,t_C,N,G)= \\
	\frac{3\gamma^2D_0G^2}{8\pi^2} \left( 4\left( N-2  \right)t_C^3 + \left( t_D-\left( N-2 \right)t_C \right)^3 \right).
	\label{Eq_NOGSEsin_libre_Ap}
\end{multline}

\end{document}